\documentclass[prl,twocolumn,showpacs,preprintnumbers,amsmath,amssymb]{revtex4}

\usepackage[dvips]{graphics}

\begin{document}

\title{Non-commuting degrees of freedom in linear quantum gravity}

\author{Chiara Marletto and Vlatko Vedral}
\affiliation{Clarendon Laboratory, University of Oxford, Parks Road, Oxford OX1 3PU, United Kingdom and\\Centre for Quantum Technologies, National University of Singapore, 3 Science Drive 2, Singapore 117543 and\\
Department of Physics, National University of Singapore, 2 Science Drive 3, Singapore 117542 and\\Fondazione ISI, Torino}

\date{\today}

\begin{abstract}
We analyse our recently proposed experiment to witness indirectly non-commuting degrees of freedom in gravity, in the light of the analogy between the electromagnetic and the gravitational field. We thereby identify the non-commuting gravitational degrees of freedom in the linear regime of Einstein's General Relativity. They are the electric-like and the magnetic-like components of the Christoffel symbol, in the weak limit of gravity. The equivalence principle can then be used to suggest further experiments of the quantum nature of gravity, exploiting acceleration only.  \end{abstract}


\maketitle                           


In the weak limit, the gravitational field of General Relativity (GR) resembles the electro-magnetic (EM) field \cite{gravitomagnetism}. In this linear regime of weak gravity, the Christoffel's symbols of GR can be expressed in a form analogous to the EM field tensor, $F_{\mu\nu}$ - \cite{Peres}. The electric-like components in the GR Christoffel tensor, $\Gamma_{\mu\nu}$, are the usual Newtonian gravity, while the magnetic-like components are those responsible for the so called inertial-induction gravitational effect (as we will see in more detail below). The latter is the gravitational version of Faraday induction and it has never been observed experimentally. The reason for its elusiveness is that the induced gravity by a moving mass is much weaker than the Newtonian gravity by the same mass.  

Recently, a different experimental scheme \cite{MAVE,MAVE2}, \cite{SOUG} has been proposed to witness non-classicality in gravity. The key idea of the experiment is that if two masses become entangled through an interaction mediated locally by the gravitational field only, then the field itself must be non-classical - in the sense that the field must have at least two non-commuting degrees of freedom. In this letter we will analyse the implications of this experiment in the weak limit of gravity, guided by the analogy with the EM field. We will follow the simple logic that if gravity is a quantum field, in the weak limit the electric-like and the magnetic-like gravitational components should obey the same commutation relations as the electric and magnetic field \cite{Peres}. A direct measurement of the non-commutativity of different components of the gravitational field is out of the experimental reach due to the weakness of the gravitational interaction itself. The proposed experiment allow this difficulty to be circumvented, by measuring the non-commutativity through the establishment of spatial entanglement.  

In this letter we will restrict attention to the quantum version of linear General Relativity, because all models of canonical quantum gravity agree on predictions in this regime. The degree of non commutativity of the electric-like and magnetic-like components of gravity can be understood by comparison with the EM case \cite{Bohr,Heisenberg,Heitler}. The uncertainty relation for the electromagnetic field confined to a region of size $L$ is $\Delta E \Delta H \geq \hbar c/L^4$, where this holds for the components of the EM field in two orthogonal directions ($H=B/\mu_0$). This degree of non-commutativity is already comparable to unity for confinements of the micron size $L\approx 1\mu$m (the Lamb shift, the Casimir effect and spontaneous emission are observable consequences of this non-commutativity). 

For gravity, the uncertainty in the metric $g$ is of the order of $\delta g\geq l_P/L$, where $l_P$ is Planck's length \cite{Regge} (see also \cite{Kiefer,Wheeler}). Since $\Gamma \propto \partial g$, the uncertainty in the components of the GR field is of the order of $\Delta \Gamma \geq l_P/L^2$. We therefore expect the electric-like and the magnetic-like components of $\Gamma$ to obey the following commutation relation in linear quantum gravity:
\begin{equation}
\Delta \Gamma_1 \Delta \Gamma_2  \geq \frac{l^2_P}{L^4} = \frac{\hbar G}{c^3 L^4} \; ,
\end{equation}
which is a relation already argued for by Peres and Rosen \cite{Peres}. Note, however, that this is of the order of unity only if $L\approx \sqrt{l_P}$ which is smaller than the nuclear dimensions. Hence, we expect the effects of this non-commutativity in quantum gravity to be much weaker. In fact, some authors have claimed that none of the gravitational phenomena that mirror the EM case (such as spontaneous emission of gravitons, gravitational Lamb shift or gravitational Casimir effect) will ever be experimentally accessible. 

Our entanglement-based witness of quantum effects in gravity in \cite{MAVE} was precisely proposed to rebut those claims. We now revisit our experimental proposal in the light of the above commutation relations in order to identify the physical meaning of the two components of $\Gamma$ that would not commute, should the experiment lead to the predicted entanglement between the two interfering masses. 

It is helpful to first review the analogue EM scenario. Consider two charges that are moving parallel to each other. The force due to their electric fields will be repulsive, while magnetically, they will attract each other. The magnetic attraction is induced by the field of the moving charges (in a completely symmetric fashion in this case). The same is the case for gravity, other than the fact that the gravitational forces are always attractive. Imagine two masses $M$ moving parallel to one another at a distance $r$ as in our double interference proposal. 
Here, the electric-like gravity is given by
\begin{equation}
E_G = \frac{1}{\epsilon_G}\frac{M}{r^2} \; ,
\end{equation}  
where we use $\epsilon_G = 1/4\pi G$ in order to emphasize the EM analogy, while the magnetic-like component is
\begin{equation}
B_G = \mu_G \frac{I}{r} \; ,
\end{equation} 
where $I$ is the mass current density and $\mu_G = 4G/c^2$ (and, as in the EM case, $c^2 = 1/\epsilon_G\mu_G$, which is expected since the gravitational waves also propagate at the speed of light). These are the two components of gravity whose non-commutativity would be revealed by witnessing entanglement in our experimental proposal. 

Note that these two non-commuting components correspond to the two non-commuting degrees of freedom $a^{\dagger}a$ (representing the energy of the gravitational field per mode) and $a+a^{\dagger}$ (representing the quantised perturbation of the metric per mode) in the linear-gravity Hamiltonian presented in \cite{MAVE2}; both are necessary for establishing the entanglement between the masses. The linear quantum gravity, therefore, perfectly reflects the intuition based on the EM analogy with weak gravity. It should also be said that the Christoffel-based commutation relations of the gravitational field can formally be derived in this linear regime and they would fully corroborate the above formula.

The following intuitive argument relying on consistency with ordinary quantum physics shows why these two components of gravity should be complementary. The resulting forces from the orthogonal $E$ and $B$ fields are co-linear (just like the Lorentz force between two charges moving in parallel). If we want to measure them simultaneously, we need a test charge which will respond to them. From the above formulae we obtain: 
$\Delta E_G \Delta H_G \geq \frac{G\hbar}{c^3r^3 L}$, which agrees with the above $\Delta \Gamma_1 \Delta \Gamma_2$ formula when $L=r$. 

However, as we said, the gravito-electric and gravity-magnetic components are weak in comparison with EM. In our entanglement-based proposal, the two masses should be roughly a nanogram, separated by a micron \cite{MAVE} in order for entanglement to be experimentally observable. Suppose that the masses are both moving at speeds of about a million meters per second (we are assuming the highest speeds that are still non-relativistic to a good approximation). The force experienced by each mass $M$ is the gravitational Lorentz force given by $F=M(E_G + v\land B_G)$ and is the sum of the electric- and the magnetic-like components. For our experiment the electric-like force is of the order of a yoctoNewton, while the magnetic-like force is three orders of magnitude smaller. Again, it should be noted that although each could be hard to detect in practice (although, see \cite{yocto}), their non-commutativity can in principle be detected indirectly by means suggested by us and Bose et al. 

We conclude with what seems to us an interesting speculation. The equivalence principle suggests that gravity and acceleration are locally indistinguishable. One therefore wonders if the quantum nature of the gravitational field could be tested purely in accelerating frames, such as using the Sagnac interferometer \cite{Sagnac}. Suppose that an interferometer could be spun at a superposition of two different angular frequencies. The massive particle undergoing interference in this kind of a rotating frame would then presumably be entangled to the angular momentum of the interferometer. Namely, one angular frequency would produce one type of interference while the other one would produce a different one. Showing that these two degrees are entangled, together with the fact that the field acts locally, would be an alternative proof of the quantum nature of gravity, albeit without using gravity at all. 

\textit{Acknowledgments}: CM thanks the Templeton World Charity Foundation and the Eutopia Foundation. VV thanks the National Research Foundation, Prime Minister's Office, Singapore, under its Competitive Research Programme (CRP Award No. NRF- CRP14-2014-02) and administered by Centre for Quantum Technologies, National University of Singapore.This research was also supported by
grant number (FQXi-RFP-1812) from the Foundational Questions Institute and Fetzer Franklin
Fund, a donor advised fund of Silicon Valley Community Foundation.

\end{document}